\documentclass[conference]{IEEEtran}
\IEEEoverridecommandlockouts

\usepackage{cite}
\usepackage{amsmath,amssymb,amsfonts}
\usepackage{algorithmic}
\usepackage{graphicx}
\usepackage{textcomp}
\usepackage{xcolor}
\usepackage{hyperref}
\usepackage{lipsum}%
\usepackage[pscoord]{eso-pic}
\newcommand{\placetextbox}[3]{%
  \setbox0=\hbox{#3}
  \AddToShipoutPictureFG*{
    \put(\LenToUnit{#1\paperwidth},\LenToUnit{#2\paperheight}){\vtop{{\null}\makebox[0pt][c]{#3}}}%
  }%
}%

\usepackage{siunitx}

\def\BibTeX{{\rm B\kern-.05em{\sc i\kern-.025em b}\kern-.08em
    T\kern-.1667em\lower.7ex\hbox{E}\kern-.125emX}}
\begin{document}

\title{On the Suitability of Wi-Fi for Interconnecting Moving Equipment in Industrial Environments \\
\thanks{This work was partially supported by the European Union under the Italian National Recovery and Resilience Plan (NRRP) of NextGenerationEU, partnership on ``Telecommunications of the Future'' (PE00000001 - program ``RESTART'').}
}

\author{
    \IEEEauthorblockN{
    Pietro Chiavassa,
    Stefano Scanzio,
    Gianluca Cena
    }  
    \IEEEauthorblockA{National Research Council of Italy (CNR--IEIIT), Italy. }    
    Email:  \{pietrochiavassa, stefano.scanzio, gianluca.cena\}@cnr.it
}

\placetextbox{0.5}{1}{This is the author's version of an article that has been published.}
\placetextbox{0.5}{0.985}{Changes were made to this version by the publisher prior to publication.}
\placetextbox{0.5}{0.97}{The final version of record is available at \href{https://doi.org/10.1109/WFCS63373.2025.11077620}{https://doi.org/10.1109/WFCS63373.2025.11077620}}%
\placetextbox{0.5}{0.05}{Copyright (c) 2025 IEEE. Personal use is permitted.}
\placetextbox{0.5}{0.035}{For any other purposes, permission must be obtained from the IEEE by emailing pubs-permissions@ieee.org.}%

\maketitle

\begin{abstract}
To ensure an unprecedented degree of flexibility, next-generation Industry 4.0/5.0 production plants increasingly rely on mobile devices, e.g., autonomous mobile robots and wearables.
In these cases, a major requirement is getting rid of cables through the adoption of wireless networks.
To this purpose, \mbox{Wi-Fi} is currently deemed one of the most promising solutions.

Achieving reliable communications over the air for distributed real-time control applications is, however, not devoid of troubles.
In fact, bounded transmission latency must be ensured for most of the exchanged packets.
Moreover, for devices powered on batteries, energy consumption also needs to be taken into account.
In this paper, a joint simulated analysis of these aspects is carried out to quantitatively evaluate what we can practically expect from Wi-Fi technology.
\end{abstract}


\section{Introduction}
Because of their ability not to rely on cables for communication, wireless networks are being increasingly used in industrial scenarios as well.
Doing so offers an unprecedented level of flexibility in deploying equipment, especially in the brownfield (retrofitting of plants to comply with Industry 4.0 and 5.0).
Most important, unconstrained mobility is also supported, 
which helps in pursuing the expected Human-Centric shift by means of, e.g., wearable devices \cite{2015-COMST}.
High-performance solutions like IEEE 802.11 wireless local area networks (WLAN) \cite{IEEE802.11-20}, also known as \mbox{Wi-Fi},
and cellular networks (5G and, in perspective, 6G), are today the preferred options for  
applications demanding high bandwidth and short latency, like virtual and augmented reality  \cite{2020-IEEE-ACC-WiFi5G} in industrial environments, as well as for supporting the interconnection of automated guided vehicles (AGV) and autonomous mobile robots (AMR) in modern production plants and warehouses \cite{2020-JIOT-WIA}.
In particular, \mbox{Wi-Fi} use is completely free of fees, which noticeably decreases its total cost of ownership in the medium-to-long term.
Moreover, it relies on the same addressing space as Ethernet (\SI{6}{B} MAC addresses) and has a very similar maximum transfer unit, which makes interconnection between the wired backbone and wireless extensions very simple.

So far, works in the literature that deal with the use of \mbox{Wi-Fi} in industrial environments have mainly focused on reliability and determinism \cite{2023-TNSM-TSN, 2019-TII-SDMAC},
which are often considered the sole relevant aspects when interconnecting devices (sensors, actuators, and controllers) in real-time control systems.
However, even in those contexts where resources are not severely limited, sustainability is nowadays regarded as another key requirement \cite{2023-ACCESS-SS, systex_pietro}.
When a device communicates over the air, two kinds of resources must be taken into account for which a cautious use is beneficial, namely, energy and spectrum.

Small AGVs, like those used for surveillance, are generally provided with batteries of limited capacity.
Hence, although the power consumption of \mbox{Wi-Fi} modules is typically small compared to motors, it could nevertheless be non-negligible.
Imagine, e.g., a rover provided with one or more cameras (conventional, infrared, time-of-flight) and some sensors (proximity, temperature, acoustic, vibration, smoke, gasses) that continually acquire images/data while moving (perception), which are processed at the edge (to discover anomalies) before being sent to the cloud (where they are stored for post-analysis and decisions are made at runtime, including what the rover must do next). 
Any savings concerning the energy spent for communication increases the lifetime and the range of such devices (before they must navigate to the charging station), which is desirable in mission-critical contexts.
This is even more true for wearable devices~\cite{wearables_survey}, including connected exoskeletons \cite{2021-ACCESS-EXO}, whose weight must be as light as possible.

Likewise, not wasting the wireless spectrum is a primary requirement when operating in unlicensed bands, which are becoming more and more crowded.
In fact, the $\SI{2.4}{GHz}$ band suffers from very high usage, and the same holds, to a lesser extent, for the $\SI{5}{GHz}$ band.
The new $\SI{6}{GHz}$ band is expected to offer drastic improvements, but only in the short-medium term.
Devising reliable and deterministic wireless communication technologies that strive to conserve spectrum resources is today another relevant challenge \cite{2019-TWC-wired, 2024-JIOT-TSCH}.

In this paper, we propose a joint analysis that considers both the typical key performance indicators (KPI) relevant to industrial applications (end-to-end latency and packet losses) and aspects related to sustainability (energy and spectrum consumption).
We analyze how these quantities vary for a device depending on its position inside a given area, and in particular, versus its distance from an access point (AP), with the aim to define approximate models. 
Besides, we will also consider the presence of interfering devices.
This work is preliminary to the definition of algorithms for modeling \mbox{Wi-Fi} communication quality.

The paper is structured as follows: Section~\ref{sec:Background} provides background information on the Wi-Fi protocol while 
Section~\ref{sec:Contributions} describes the contribution of this paper.
Section~\ref{sec:Methodology} and Section~\ref{sec:Results} present the adopted methodology and obtained results, respectively.
Finally, Section~\ref{sec:Conclusions} proposes a discussion on the outcomes of this work and draws some conclusions.

\section{Wi-Fi Basics}
\label{sec:Background}
Unlike wired networks like Ethernet, wireless communication is by its nature unpredictable.
More specifically, the likelihood that a transmission attempt may fail (that is, when the frame is not received correctly by the intended destinations) cannot be ignored, and depends on many factors related to both the environment (signal attenuation, multipath fading) and the presence of other equipment (electromagnetic noise produced by power machinery, interference generated by nearby nodes).
To tackle this intrinsic unreliability, automatic retransmission is customarily adopted by wireless protocols, which requires the sender to set a timeout on every transmitted frame and the recipients to explicitly acknowledge any received frame.
In \mbox{Wi-Fi}, the ACKtimeout duration is defined by the IEEE 802.11 specification
and an ACK frame must be returned by the receiver within a short interframe space (SIFS),
whose duration is $\SI{10}{\mu s}$ in the $\SI{2.4}{GHz}$ band and $\SI{16}{\mu s}$ on $\SI{5}{GHz}$.

Retransmissions substantially decrease the likelihood that a message is definitely lost, 
to the point that, in theory, the packet loss ratio (PLR) seen by applications could approach zero,
as in wired networks (provided that communication between STA and AP is not prevented for a while).
On the other hand, transmission latency may increase tangibly due to the exponential backoff, which doubles the collision window on every retry.
While this is perfectly acceptable for best-effort traffic (including multimedia streaming), problems are likely to arise for time-sensitive applications, especially in industrial environments.
In that case, messages exceeding their intended deadlines are useless, and hence not dissimilar from those that went lost.
Metrics like the deadline miss ratio (DMR) can be defined, which account for both lost and late messages.

To model \mbox{Wi-Fi} communication properly, rate adaptation (RA) mechanisms must also be taken into account.
In fact, they are customarily supported in every real implementation (typically in the driver, exploiting specific hardware 
 support made available by the network board).
A very popular RA mechanism is Minstrel~\cite{minstrel}, which adaptively changes the modulation and coding scheme (MCS), therefore affecting the data rate, to optimize performance.
It relies on both exploration and exploitation (through probe and normal frames, respectively), contextually evaluating the quality of service (QoS) for all available MCSs.
Theoretically modeling the behavior of Wi-Fi with RA is a very complex task, and, for this reason, simulation is often preferred for analyzing communication quality.

In the following, industrial control systems are considered where devices are connected over the air and exchange small-sized process data (in the order of one hundred bytes) either cyclically or sporadically.
For this reason, features aimed at increasing throughput, like multiple streams and channel bonding (introduced with Wi-Fi 4 and 5), typically make little sense.
For example, if the payload is increased by $\SI{27}{B}$ when transmitting with 
orthogonal frequency-division multiplexing (OFDM)
at $\SI{54}{Mb/s}$, the air time just increases by one symbol ($\SI{4}{\mu s}$).
Instead, it is essential that, with a very high likelihood (e.g., $\geq 99.9\%$), these messages are delivered to their destination within the intended deadlines, which are typically in the order of a few milliseconds.
Concerning frame aggregation, it is worth pointing out that, in control systems, any instance of a periodic process data stream should be typically delivered before the next instance is generated, which makes this mechanism not as useful as for (bursty) elastic traffic (often encountered in consumers and office applications).
Conversely, \mbox{Wi-Fi} 6 and 7 define some new interesting mechanisms, like trigger frames and multi-link operation (MLO), which could offer sensible advantages for industrial applications.
However, they are not widely spread (few commercial devices exist that fully support these features) and the related analysis is left as future work.

\mbox{Wi-Fi} behavior is affected by both geometric and operational aspects, i.e., where nodes and interferers are located (which affects signal attenuation) and the traffic pattern they generate
(which affects interference).
Other phenomena will also be explicitly analyzed in this paper, like the hidden node (which may impair communication more than people usually expect).
Although it can be tackled in theory by means of standard request to send / clear to send (RTS/CTS) primitives, we will not rely on them.
Again, this is because we are considering small-sized messages, whose duration is similar to RTS and CTS control frames.
Therefore, from a probabilistic point of view, protecting the related transmissions by prepending an RTS/CTS exchange is pointless.

\section{Contribution}
\label{sec:Contributions}
This paper analyzes the communication quality observed by a wireless station (STA) associated to an AP in an infrastructure \mbox{Wi-Fi} network (the typical arrangement in a real-world setup), versus both their relative positions and the presence and placement of interferers operating in the same frequency range.
We refer to that STA with the term \textit{STA under test} (SUT), to distinguish it from \textit{interfering} STAs (INT).

The main goal of this analysis is to study and quantify the effects and phenomena that negatively affect Wi-Fi behavior.
In addition to the traditional performance metrics about reliability and timeliness, sustainability aspects such as energy and spectrum consumption are also considered.
The results obtained in this work are meant to constitute the starting point for the development of simplified models to (approximately) describe (and optimize) the QoS offered by industrial \mbox{Wi-Fi} networks.
In turn, these models can be employed in network digital twins (NDT) aimed at optimizing at runtime the operation of real-time applications where devices (e.g., fleets of AMRs) are connected over the air, including seamless handover management, and redundant transmission strategies \cite{10144228}.
Although the use of NDTs in 5G/6G networks has been already envisaged for a few years \cite{2023-ACCESS-NDT}, what we seek are non-resource-hungry algorithms that can run on inexpensive Wi-Fi equipment \cite{2024-WFCS-CAVALC, 2024-WFCS-SCANZIO}.

\section{Methodology}
\label{sec:Methodology}

In the experiments we carried out, both the SUT and INT operate in the same frequency range and are connected to the same AP.
Different scenarios were considered, which differ by the relative positions of the wireless nodes in a one-dimensional space (i.e., over a line).
This approach is meant to lower complexity, while still providing results that can be generalized to many two- and three-dimensional scenarios.
Since a typical shop floor is considered, the presence of obstacles, such as walls, is also neglected, only focusing on attenuation and interference in the most general case.
Network behavior is analyzed through simulation via the ns-3 library.

\subsection{Network configurations}
\begin{figure}[b]
    \centerline{\includegraphics[width=0.95\columnwidth]{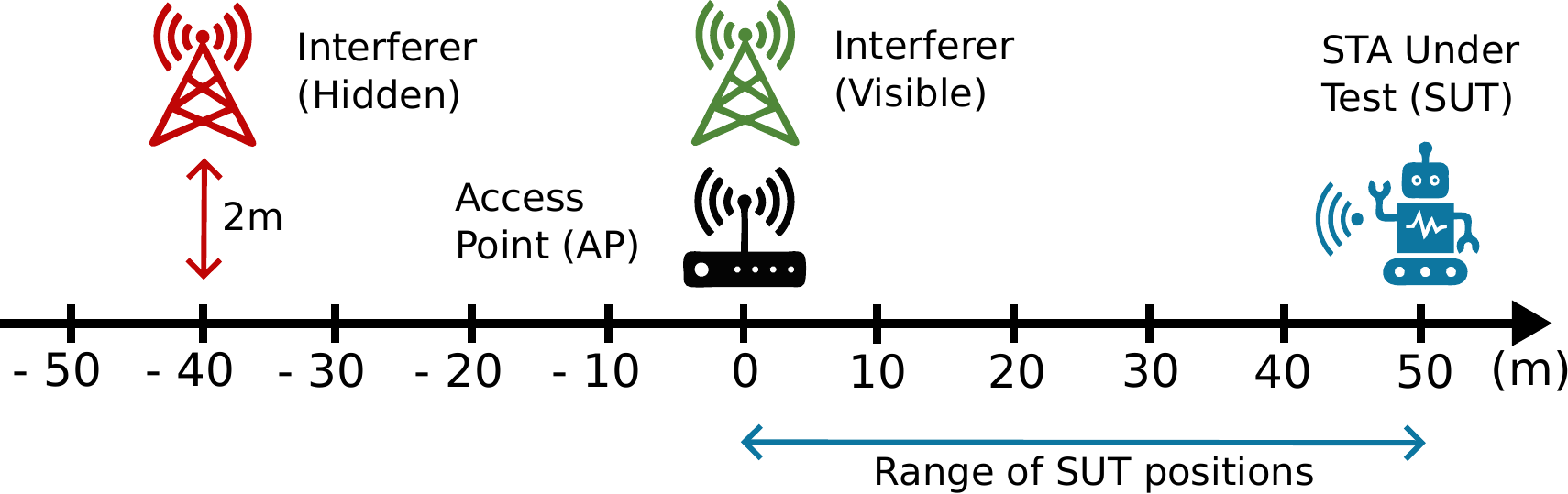}}
    \caption{Physical placement of devices (one-dimensional map).}
    \label{fig:Map}
\end{figure}

Three main classes of network configurations are analyzed and schematized in Fig.~\ref{fig:Map}.
In all these configurations, the AP is positioned at the origin of a one-dimensional Cartesian coordinate system.
The SUT, instead, is initially located at a distance $D_\mathrm{S}=\SI{1}{m}$ from the AP (on its right side, i.e., in the direction of positive values).
Then, it is moved away from the AP in steps of $\SI{1}{m}$, until $D_\mathrm{S}$ reaches $\SI{50}{m}$.
After this point, more precisely at $D_\mathrm{max}=\SI{51.45}{m}$, the SUT is no longer able to receive beacons from the AP, and therefore no frames can be exchanged between the two.
This is a consequence of the specific propagation model we selected in ns-3.

A separate (and independent) simulation is carried out for every given position of the SUT, to characterize the QoS it experiences in different places in stationary conditions (after all transient phenomena have settled).
Please note that the results we obtained should not be regarded as if the SUT was moving at constant speed in a straight line.
They could be used to approximate the communication quality in case of SUT mobility, but only if its speed is low enough.
Conversely, if the SUT moves too quickly, optimizations performed by the RA mechanism may be unable to converge to the performance achieved in the absence of mobility.
For the sake of simplicity, handover is also not considered. That is, no other APs are available to which the SUT can re-associate when it moves too far from the origin.
The three configurations are as follows:

\paragraph{{NO\_INT configuration}}
The first network configuration considers the case where no interfering devices are present.
Results can be seen as a baseline, to better highlight the effects of interferers in subsequent experiments.
More in detail, this configuration allows us to isolate the effects of signal attenuation and RA on the perceived QoS.
In fact, when $D_\mathrm{S}$ grows and signal attenuation increases, transmission failures will cause RA to lower the data rate, making the duration of each single attempt longer.
This does not necessarily result in a latency increase, as the mean number of attempts usually diminishes since more robust modulations are used.

\paragraph{{VISIBLE configuration}}
The second network configuration considers the presence of a single interfering STA located at $D_\mathrm{I}=\SI{0}{m}$, that is, in the same place as the AP.
Past experience taught us that disturbance in real industrial plants is often due to other Wi-Fi devices.
To avoid simulating a potentially problematic scenario (where two antennas occupy exactly the same location), the interferer was displaced by a small offset ($\SI{2}{m}$) in the perpendicular direction.
The interferer is associated with the AP and continuously injects traffic on air.
In this setup, the interferer is always visible by the SUT across its entire range of movement, meaning that the SUT is always able to sense the interfering traffic.
This condition allows the STAs to effectively exploit carrier sensing techniques to postpone transmissions when the channel is busy, hence preventing many collisions.

\paragraph{{HIDDEN configuration}}
In the third network configuration, the interfering STA is placed at $D_\mathrm{I}=\SI{-40}{m}$ (i.e., on the opposite side with respect to the SUT).
Although not strictly necessary, the $\SI{2}{m}$ perpendicular offset is maintained for consistency.
In this scenario, when $D_\mathrm{S}> \SI{11}{m}$
(i.e., the SUT and the interferer are more than $\SI{51.45}{m}$ apart),
they are no longer able to detect each other's traffic
(according to the ns-3 propagation model).
This condition, commonly referred to as the hidden-node problem, severely impairs collision avoidance, making the MAC layer operate sub-optimally and leading to much more frequent collisions between uplink packets (directed to the AP).

\subsection{Simulation setup}

For the reasons stated above concerning the characteristics of periodic real-time traffic in industrial distributed control applications, the IEEE 802.11a standard (which relies on OFDM) is selected for all simulations.
Devices were set to operate in the $\SI{5}{GHz}$ frequency band on channel 44, adopting the standard channel width of $\SI{20}{MHz}$.

In ns-3, the PHY layer is simulated via the \textit{SpectrumWifiPhy} class, using the \textit{SpectrumChannel} class to describe channel conditions.
This class is configured using the \textit{ConstantSpeedPropagationDelayModel} and \textit{LogDistancePropagationLossModel} objects, which characterize the signal propagation speed and path loss, respectively.
With these models, the propagation speed $s_{prop}=\SI{2.99792e+08}{m/s}$ (ns-3 default) is kept constant, while the path loss is evaluated as
\begin{equation}
    L = L_0 + 10 \cdot n \cdot \log_{10} \left( \frac{D}{D_0} \right),
\end{equation}
where $L$ is the path loss (expressed in $\SI{}{dB}$) at distance $D$, $L_0=\SI{46.6777}{dB}$ is the path loss at the reference distance $D_0=\SI{1}{m}$, and $n=3$ is the path loss distance exponent (ns-3 default values).
\textit{SpectrumWifiPhy} also specifies the transmission power of the Wi-Fi transmitter, $P_\mathrm{tx}=\SI{16.0206}{dBm} (\sim\SI{40}{mW})$,
and the minimum power threshold for preamble detection $P_\mathrm{min}=\SI{-82}{dBm}$ (ns-3 default values).
With this configuration, the maximum allowed transmission distance is $D_\mathrm{max}=\SI{51.45}{m}$.

In all the tested network configurations, the SUT, the interfering STA, and the AP are set up to use Minstrel as the RA algorithm.
In addition, the packet size threshold to use RTS/CTS is suitably increased for all STAs in order to disable this feature.
The maximum number of allowed packet retransmissions (retry limit) for the SUT is $R_\mathrm{L}=13$.

The SUT and the interfering STA run two different applications that generate traffic on the Wi-Fi network.
Both applications send UDP packets to a server installed on the AP.
The SUT application generates a new packet every $\SI{0.5}{s}$ with a $\SI{22}{B}$ payload, which is transmitted on air as a single IEEE 802.11 PSDU that includes $\SI{86}{B}$ in total.
This type of traffic resembles the exchange of control messages in a typical industrial application.
The UDP payload is composed of a $\SI{12}{B}$ custom header that contains a timestamp taken when the packet was sent and a sequence number that uniquely identifies it, followed by random data.
The timestamp is used to evaluate transmission latencies, while the sequence number serves to track every single packet across the simulation.

The application running on the interferer generates instead UDP bursty traffic.
The idle time between subsequent bursts is selected randomly according to an exponential distribution with a mean of $\SI{250}{\mu s}$, and is clipped to $\SI{10}{s}$.
The number of packets in every burst is also selected from a random exponential distribution, with a mean of $100$ packets and upper-bounded to $500$.
Packets in each burst are spaced by $\SI{500}{\mu s}$ and their payload includes $\SI{1472}{B}$.
This kind of traffic is used to mimic general-purpose applications that execute bandwidth-intensive operations (file transfer, multimedia streaming, etc.).

The simulation is run for $\SI{30000}{s}$, during which the SUT generates about $\SI{60000}{}$ packets, which are sent using confirmed unicast transmission services.
The ARP tables of all nodes are pre-populated to prevent delays at the beginning of the simulation.
The only exception is the initial association procedure between the STAs and the AP.
To account for this, both client and server applications are started at simulation time $T_\mathrm{s} = \SI{1}{s}$, so that they find a fully operating link and no additional delays are experienced.

\subsection{Evaluation metrics}
Evaluation is carried out starting from the output of every simulation, which consists of a log of all the transmission attempts for the packets generated by the SUT.
For every packet, the overall latency and the transmission outcome (either \textit{acked} or \textit{dropped}) are recorded.
Instead, every transmission attempt is characterized by the transmission power $P_\mathrm{tx}$ (always $\sim\SI{40}{mW}$, in our case), data rate $r_i \in \{6, 9, 12, 18, 24, 36, 48, \SI{54}{Mb/s} \}$, and transmission time ${T}_\mathrm{tx}$,
which refers to the time needed to transmit the PSDU and the related preamble.
Since we chose a fixed PSDU size, this time only depends on the selected data rate.

The analysis we made concerns not only the QoS, intended as reliability and timeliness, but also resource utilization, in terms of energy and spectrum consumption.
In particular, \textit{reliability} can be measured as the fraction of packets that are delivered successfully, possibly exploiting one or more retries, 
whereas \textit{timeliness} can be quantified by analyzing the latency of packet delivery.
Concerning \textit{spectrum} consumption, it depends linearly on the overall airtime of packets, which is obtained by summing the duration of all the related transmission attempts (including retries).
In turn, the duration of a single attempt (DATA/ACK frame exchange) consists of a fixed part (that accounts for preambles and interframe gaps) plus a contribution inversely proportional to the data rate.
Since the transmission power is constant, also \textit{energy} consumption is related linearly to the overall airtime.

Given these considerations, the following metrics are selected and evaluated for every position of the SUT in all the network configurations:
\begin{itemize}
    \item Packet loss ratio (PLR): permits to determine link reliability, expressed as $1-$PLR.
    
    \item Average value $\mu_d$ and standard deviation $\sigma_d$ of latency: we found that latency is not normally distributed, so $\sigma_d$ is not significant and is excluded from the analysis.
    
    \item Minimum latency $d_{\min}$: this is the best case, when no collisions occur and the packet is delivered immediately.
    
    \item $99$ and $99.9$ latency percentiles  ($d_\mathrm{p99}$ and $d_\mathrm{p99.9}$): they serve to characterize soft real-time performance (deadlines must be met by the majority of packets, not necessarily by all of them).
    
    \item Average number  $\mu_a \in [1,1+R_\mathrm{L}]$ of transmission attempts on air (initial plus retries).
    
    \item Average data rate $\mu_r$ with which transmission attempts are performed (as per Minstrel operation).
    
    \item Data rate selection frequency $f_{r_i}$: overall number of transmission attempts performed with data rate $r_i$.

    \item Relative success frequency $s_{r_i}$ of frame transmission attempts at data rate $r_i$.
    
    \item Mean power consumption $\mu_{P}$ related to data transmission (PSDU+preamble): computed by summing the energy contribution of every transmission attempt (equal to $P_\mathrm{tx} \cdot {T}_\mathrm{tx}$) and then dividing by the entire simulation time.

\end{itemize}
Metrics related to the latency and the number of transmission attempts are computed considering only delivered packets, while the rest also include those that are dropped.
Latency is evaluated as the time elapsing between packet generation and the reception of the related ACK, using the common time base of the simulator to obtain exact latency measurements.

\section{Results}
\label{sec:Results}
The results of the simulation campaigns conducted for the different network configurations are reported below.

\subsection{No Interfering Devices}

The first simulation campaign was carried out in the absence of any source of interference (NO\_INT configuration) and is meant to study the relationship between the position of a STA and the QoS it receives.
In particular, it was aimed at analyzing the effect of attenuation on signal propagation, as well as the benefits achieved by RA. 
For this reason, it will be taken as the baseline against which other scenarios will be checked.
Interestingly, no packets went lost, which means that $\mathrm{PLR}=0$.
This is unsurprising, as the only source of interference are the beacons sent by the AP.
While transmission attempts of the SUT may occasionally fail, a few retries are typically enough to ensure correct packet delivery.

\begin{figure}[t]
    \centerline{\includegraphics[width=0.9\columnwidth]{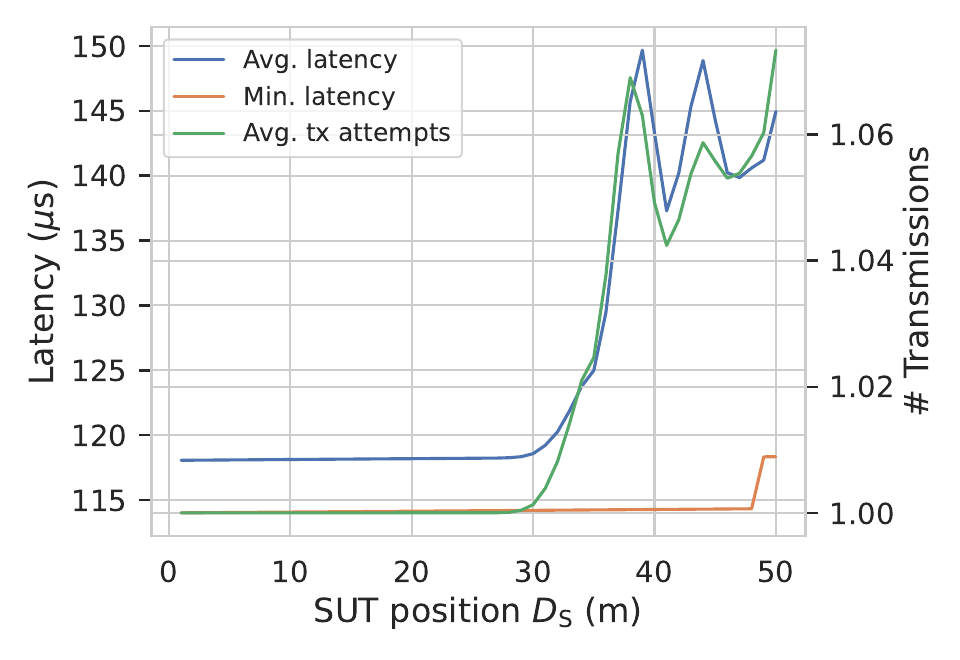}}
    \vspace{-0.4cm}
    \caption{Latency ($\mu_\mathrm{d}$, $d_{\min}$) and tx attempts ($\mu_\mathrm{a}$) vs. $D_\mathrm{S}$ (NO\_INT).}
    \vspace{-0.2cm}
    \label{fig:AvgLat}
\end{figure}

Fig.~\ref{fig:AvgLat} reports the average $\mu_d$ and minimum $d_{\min}$ transmission latency of a packet as a function of the distance $D_\mathrm{S}$, together with the average number $\mu_a$ of transmission attempts.
As can be seen, as long as $D_\mathrm{S}$ remains below $\SI{28}{m}$ $\mu_d$ does not vary significantly and is about $\SI{118}{\mu s}$.
This is slightly higher than $d_{\min}$ (about $\SI{114}{\mu s}$), which represents the best case (when the highest data rate is used and no retries are needed).
In this inner region, retransmissions are rarely encountered.
It is also interesting to see that $d_{\min}$ increases (very slightly, indeed) with distance, due to two-way signal propagation delays (about $\SI{6.7}{ns/m}$).

Moving the SUT farther away from the AP increases the failure probability for attempts, which means that more retries are performed on average per packet.
Average latency $\mu_d$ is tangibly affected by the increased mean number $\mu_a$ of attempts, as also witnessed by the strong correlation between these quantities (Pearson correlation $0.982$).
By analyzing the plots in detail, $\mu_d$ reaches a maximum of $\SI{150}{\mu s}$ when $D_\mathrm{S}=\SI{39}{m}$, to then start an alternating trend where it decreases and subsequently increases.
This is due to the combined effect of the wireless channel behavior for every given data rate, 
where the failure probability features a steep increase when the distance exceeds a specific threshold,
and Minstrel, which dynamically starts selecting lower data rates more frequently to achieve higher robustness against disturbance.

\begin{figure}[t]
    \centerline{\includegraphics[width=0.9\columnwidth]{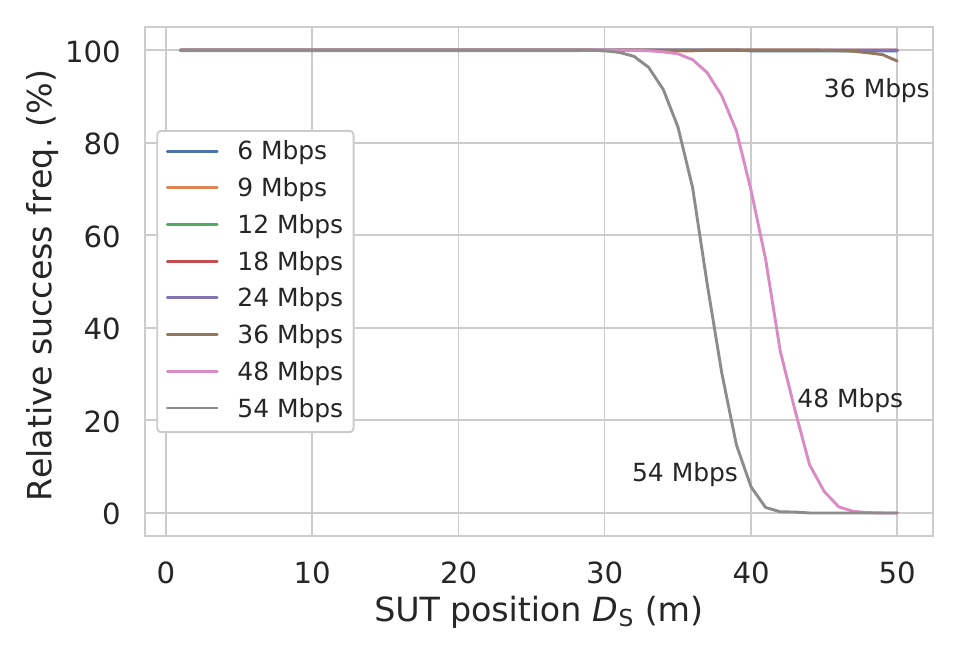}}
    \vspace{-0.4cm}
    \caption{Relative success frequency $s_{r_i}$ of each data rate $r_i$ vs. $D_\mathrm{S}$ (NO\_INT).}
    \vspace{-0.2cm}
    \label{fig:RateSucc}
\end{figure}

\begin{figure}[b]
    \vspace{-0.4cm}
    \centerline{\includegraphics[width=0.9\columnwidth]{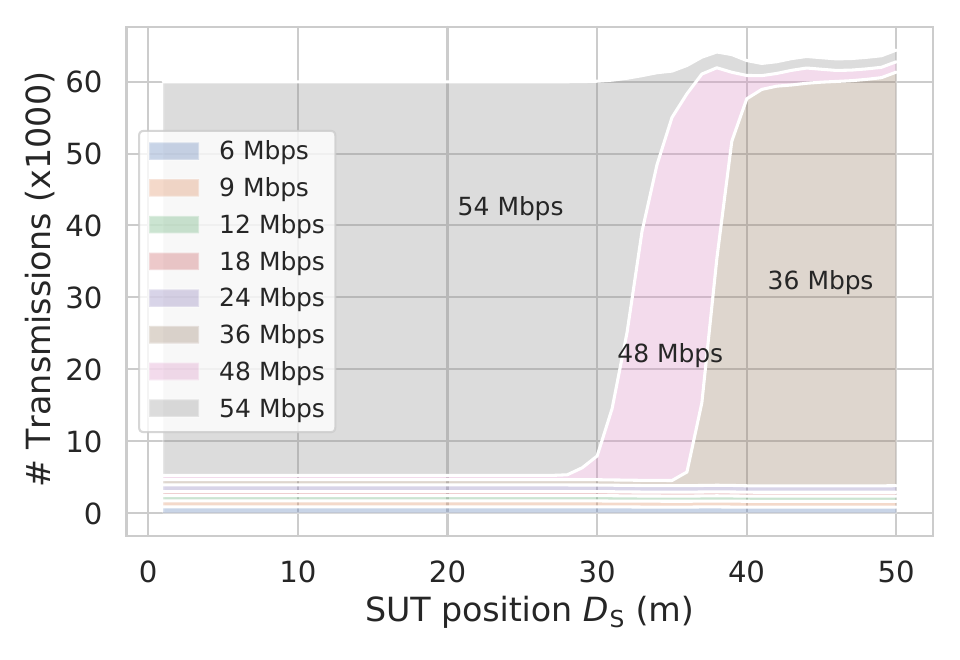}}
    \vspace{-0.4cm}
    \caption{Selection frequency $f_{r_i}$ of data rate $r_i$ vs. $D_\mathrm{S}$ (NO\_INT).}
    \label{fig:RateDist}
\end{figure}

RA effects can be better understood by observing Fig.~\ref{fig:RateSucc} and Fig.~\ref{fig:RateDist}, which depict the relative success frequency $s_{r_i}$ for transmission attempts performed with any given data rate $r_i$ and the frequency $f_{r_i}$ with which data rate $r_i$ is selected by Minstrel, respectively.
When $D_\mathrm{S}$ exceeds $\SI{28}{m}$, the success probability of the highest transmission rate ($r_{\max} =\SI{54}{Mb/s}$) starts decreasing, so Minstrel begins to prefer $\SI{48}{Mb/s}$ transmissions.
It is worth noting that, since we used quite short frames (the PSDU size is $\SI{86}{B}$), the duration ${T}_\mathrm{tx}$ of the data frame does not change in these two cases and is equal to $\SI{36}{\mu s}$ (training/signal fields plus $4$ OFDM symbols).
However, around $\SI{31}{m}$, also $\SI{48}{Mb/s}$ transmissions start to fail, which leads Minstrel to increase the usage of $\SI{36}{Mb/s}$ transmissions.

The first local maximum for $D_\mathrm{S}$ is reached at $\SI{39}{m}$, due to the high failure probability of $\SI{54}{}$ and $\SI{48}{Mb/s}$ transmissions, which are still frequently used.
By also reducing the number of $\SI{48}{Mb/s}$ transmissions, a local minimum is reached at $\SI{41}{m}$.
When the distance is increased further, Minstrel keeps the frequency with which $\SI{54}{}$ and $\SI{48}{Mb/s}$ data rates are selected almost constant, despite their failure probability still increases, which leads to a second peak when $D_\mathrm{S} = \SI{44}{m}$.
This is again faced by Minstrel, which slightly increases the usage of data rate $\SI{36}{Mb/s}$ at the expense of $\SI{48}{Mb/s}$, leading to another local minimum when $D_\mathrm{S} = \SI{47}{m}$.
Thereafter, latency increases again because also $\SI{36}{Mb/s}$ transmissions start to fail consistently.

The above example describes the operations of Minstrel from a statistical viewpoint that involves transmission data rates and outcomes of attempts, and explains why this mechanism is effective in dealing with transmission failures due to signal attenuation when a STA is moving around (a theoretical analysis is likely possible, but much more complex).
It also shows why, in some cases, reducing the data rate (counterintuitively) improves responsiveness:
attempts are, on average, longer, but their mean number may decrease more tangibly.

It is worth stressing that the Minstrel algorithm is unable to provide the best theoretical performance. 
The reason is that, to adapt to changes in channel conditions, it occasionally performs some transmissions at data rates other than the optimum for the given distance (exploration vs. exploitation).
Doing so results in a behavior that is not as good as what could be theoretically achieved 
if the SUT knew: a) its distance $D_\mathrm{S}$ from the AP, and, b)~that the channel is stationary. In that case, it could set the data rate to the corresponding optimal value, making latency monotonically increase with distance.
Clearly, these assumptions are not realistic.

If circular symmetry is exploited to analyze a more realistic 2D environment (e.g., a shop floor) starting from the results of our experiments (by setting the radius equal to $D_\mathrm{S}$),
we can see that the coverage region of the AP is split into two parts:
an inner circle, where no advantages are obtained by Minstrel, 
and an outer ring (which is more than twice as large as the former), where tangible benefits are gained.

\begin{figure}[t]
    \centerline{\includegraphics[width=0.9\columnwidth]{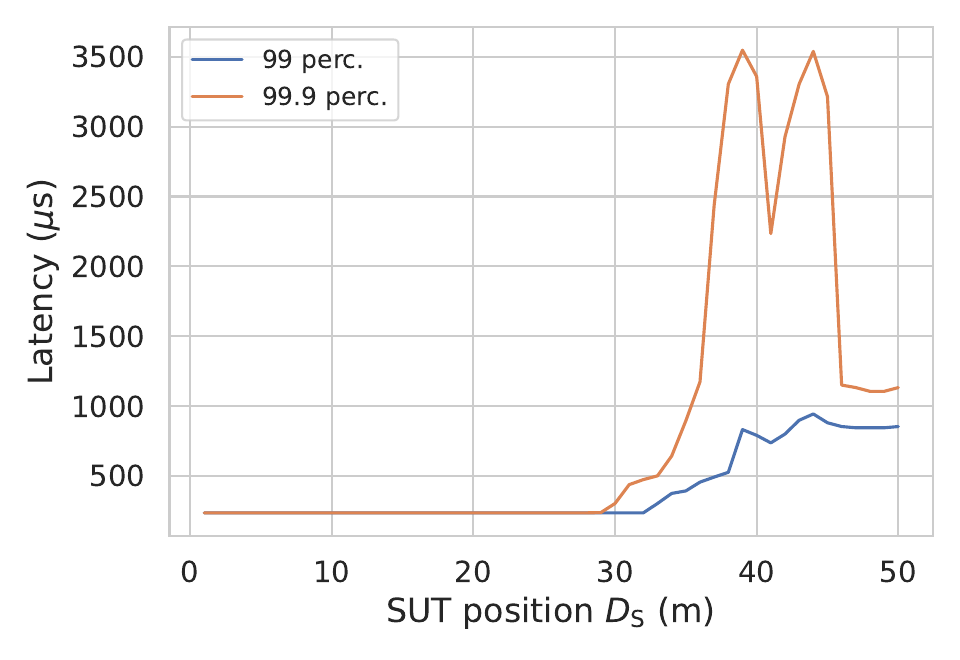}}
    \vspace{-0.4cm}
    \caption{Latency percentiles $d_\mathrm{p99}$ and $d_\mathrm{p99.9}$ vs. $D_\mathrm{S}$ (NO\_INT).}
    \vspace{-0.3cm}
    \label{fig:PercLat}
\end{figure}

Concerning Wi-Fi suitability for soft real-time systems, Fig.~\ref{fig:PercLat} shows two high percentiles, namely, $99$ and $99.9$.
What is important to observe here is that, even in mildly disturbed environments that do not include any specific source of interference, frame losses due to the background noise may lead to multiple retransmissions.
Hence, one packet over one thousand may experience a latency as high as $\SI{3.5}{ms}$.

\begin{figure}[t]
    \centerline{\includegraphics[width=0.9\columnwidth]{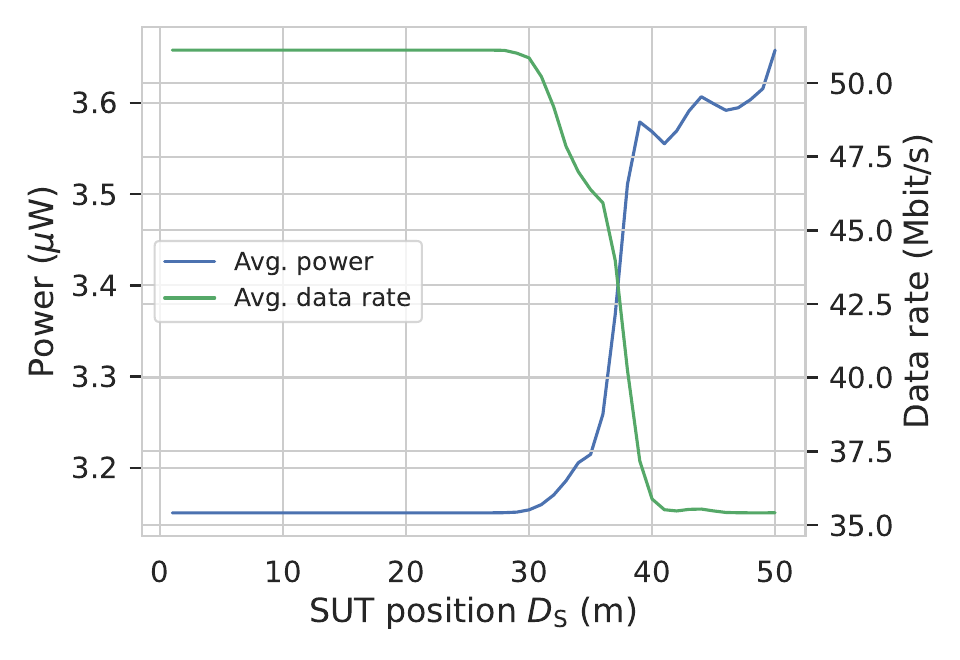}}
    \vspace{-0.4cm}
    \caption{Power consumption $\mu_{P}$ and data rate $\mu_{r}$ vs. $D_\mathrm{S}$ (NO\_INT).}
    \vspace{-0.3cm}
    \label{fig:AvgPower}
\end{figure}

Fig.~\ref{fig:AvgPower} depicts the average data rate $\mu_{r}$ used for transmissions by the SUT and its average power consumption $\mu_{P}$ versus the distance $D_\mathrm{S}$.
Power consumption depends on the number of retries (at least one attempt is always performed) and the average transmission time of frames.
For packets of fixed size (as in our experiments), the duration of the frame only depends on the data rate, which determines the number of transmitted OFDM symbols ($\SI{4}{\mu s}$ each).
Unlike latency, the trend of the power consumption for increasing distances is almost monotonic, suggesting that this quantity is affected less by the average number of attempts $\mu_{a}$.
In fact, power consumption $\mu_{P}$ grows linearly with retries, while latency $\mu_{d}$ also includes the contribution of backoffs, whose duration increases exponentially with retries.

\subsection{Single Interfering Device}
The VISIBLE and HIDDEN network configurations serve to analyze what happens to a Wi-Fi link in the presence of an interfering STA.
In the experiment, the interferer is associated to the same AP as the SUT and injects bursty traffic on air.
In the first configuration it is located near the AP ($D_\mathrm{I}=\SI{0}{m}$), while in the second it is positioned at $D_\mathrm{I}=\SI{-40}{m}$ (i.e., on the opposite side of the AP with respect to the SUT).

While the VISIBLE configuration provides valuable results about the effects of interference, 
the HIDDEN configuration is particularly interesting since it highlights the hidden-node effect.
In our ns-3 simulated environment, this happens when the SUT is more than $\SI{11}{m}$ away from (and on the right of) the AP ($D_\mathrm{S} > \SI{11}{ms}$, which implies that $D_\mathrm{S} - D_\mathrm{I} > \SI{51.45}{m}$).
In these conditions, the two STAs (SUT and INT) are unable to receive each other's frames; hence, carrier sense no longer provides proper input for collision avoidance.
As a consequence, the MAC layer shows quite a bad behavior.

\begin{figure}[b]
    \centerline{\includegraphics[width=0.9\columnwidth]{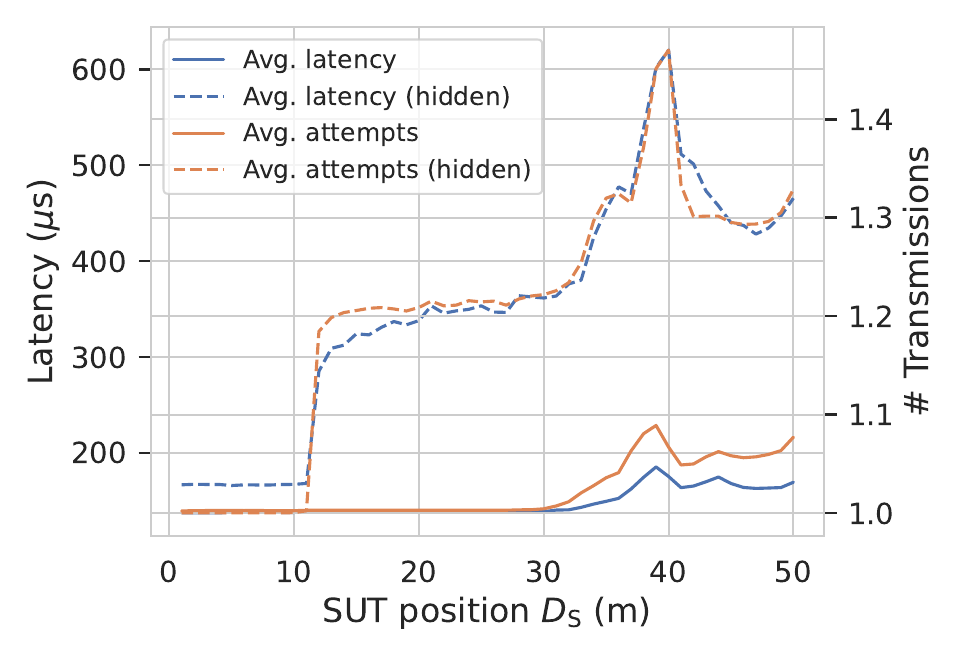}}
    \vspace{-0.4cm}
    \caption{Latency $\mu_{d}$ and tx attempts $\mu_{a}$ vs. $D_\mathrm{S}$ (VISIBLE and HIDDEN).}
    \label{fig:AvgLatInt}
\end{figure}

Fig.~\ref{fig:AvgLatInt} depicts the average latency and the average number of transmission attempts per packet for both configurations.
In the VISIBLE configuration, the trends of the two quantities are very similar to the case without interferers.
However, higher values are obtained for both metrics, due to the fact that the SUT may find the channel busy due to INT transmissions, or the two may collide, which implies that retransmissions are made after a random backoff (which, in turn, might collide again).
Latency is almost constant for $D_\mathrm{S} \le \SI{30}{m}$, even if it is not as stable as the NO\_INT case
due to the presence of the interferer 
(which generates random traffic),
while, at higher distances, the operations of Minstrel coupled with the increased failure probability for higher transmission speeds still produce local minima and maxima.

\begin{figure}[b]
    \centerline{\includegraphics[width=0.9\columnwidth]{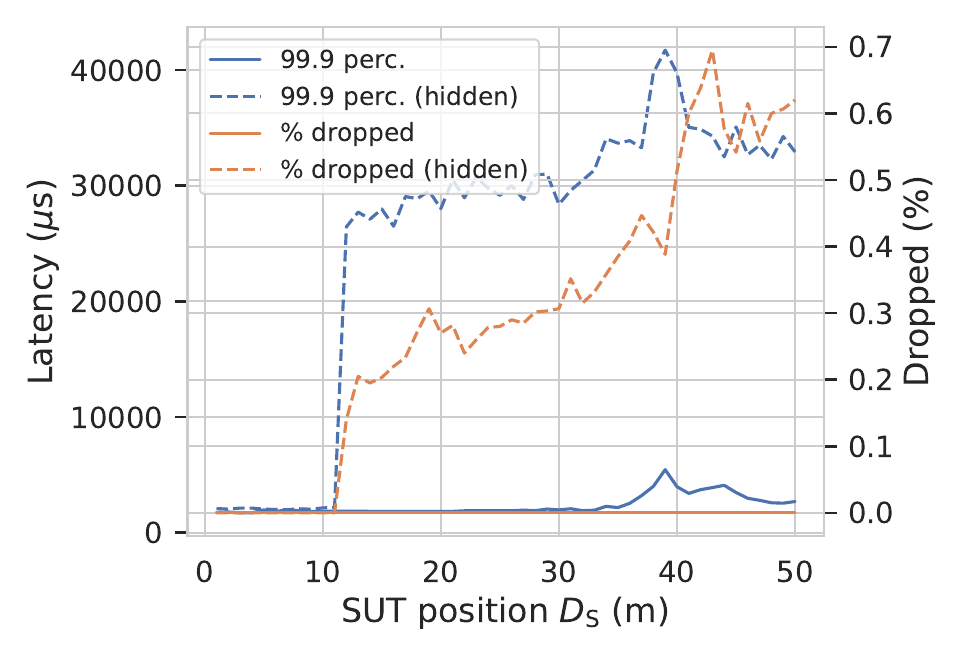}}
    \vspace{-0.4cm}
    \caption{Latency percentiles $d_\mathrm{p99}$, $d_\mathrm{p99.9}$, and PLR (\%) vs. $D_\mathrm{S}$ (VISIBLE and HIDDEN).}
    \label{fig:PercLatInt}
\end{figure}

As expected, for the HIDDEN configuration there is an abrupt increase of the average latency and the number of transmissions for $D_\mathrm{S}>\SI{11}{m}$ due to the hidden-node effect.
For $D_\mathrm{S}>\SI{31}{m}$, Minstrel effects are still relevant, leading to a peak of the mean latency of $\SI{620}{\mu s}$ when $D_\mathrm{S}=\SI{40}{m}$.
Another quite interesting aspect is the higher latency w.r.t. the VISIBLE configuration for $D_\mathrm{S} \le \SI{11}{m}$ (in which case the SUT and INT can see each other and there is no hidden node).
This is due to the fact that, in the HIDDEN configuration, the distance $\lvert D_\mathrm{I} \rvert$ of the interfering STA from the AP is quite large, and so a non-negligible number of retransmissions occur for it due to path loss.
This amplifies the interfering traffic on the channel (and its occupancy), which increases the average latency $\mu_{d}$ for the SUT.
Curiously, this growth does not affect the average number of transmission attempts $\mu_{a}$ of the SUT, as it can sense the interfering traffic and prevent most of the collisions. 

For both the VISIBLE and HIDDEN configurations, average latency and number of transmission attempts are still strongly correlated (Pearson correlation $0.983$ and $0.987$, respectively), showing that retransmissions still have a big impact on latency.

Fig.~\ref{fig:PercLatInt} shows the $99.9$ percentile of latency ($d_\mathrm{p99.9}$), together with the fraction of dropped packets.
As can be seen, packets are dropped only in the presence of the hidden node, in which
case up to $0.7\%$ of the packets may be dropped (i.e., they are not delivered).
The hidden node also has a detrimental effect on $d_\mathrm{p99.9}$, which reaches $\SI{41.7}{ms}$ at $D_\mathrm{S}=\SI{39}{m}$.
Again, we stress that, from the point of view of real-time control applications, exceeding a deadline is the same as losing a packet.
For comparison, in the VISIBLE configuration, the highest value of $d_\mathrm{p99.9}$ is $\SI{5.4}{ms}$, which is about one order of magnitude smaller.
In both cases, much worse results are likely obtained when the number of interferers is increased, since they may also collide with each other.
This will be the subject of future investigations.

\begin{figure}[b]
    \vspace{-0.3cm}
    \centerline{\includegraphics[width=0.9\columnwidth]{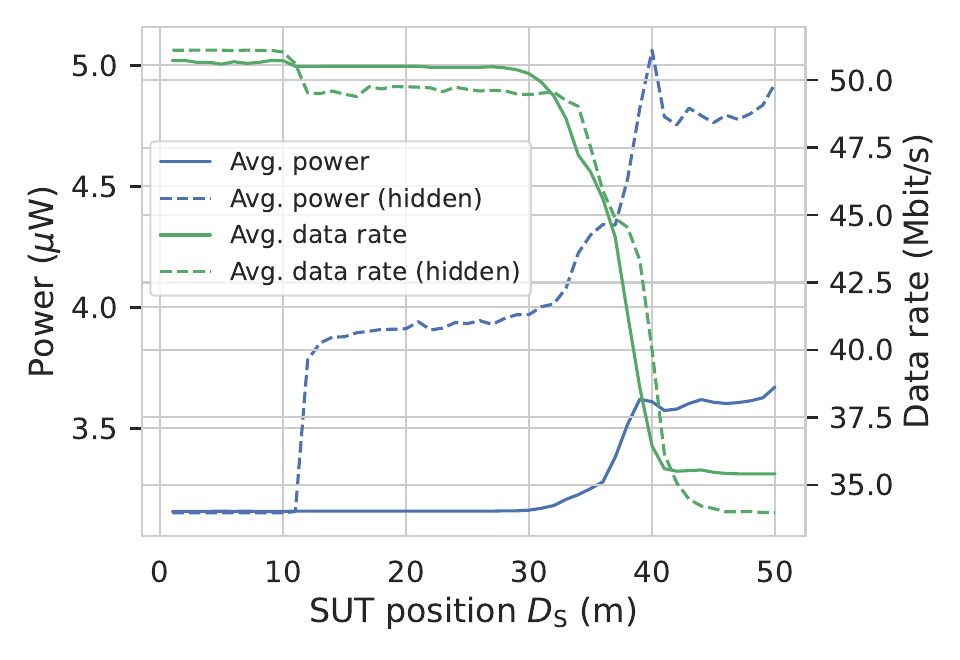}}
    \vspace{-0.4cm}
    \caption{Power consumption $\mu_{P}$ and data rate $\mu_{r}$ vs. $D_\mathrm{S}$ (VISIBLE and HIDDEN).}
    \label{fig:AvgPowerInt}
\end{figure}

Finally, Fig.~\ref{fig:AvgPowerInt} shows the average power consumption $\mu_{P}$ and the average data rate $\mu_{r}$ for the SUT.
For what concerns power consumption, there is almost no difference with respect to the baseline when adding the interferer in $D_\mathrm{I}=\SI{0}{m}$ (VISIBLE configuration).
The effect of the hidden node in the HIDDEN configuration, instead, is clearly visible. 
This phenomenon causes a sudden surge of the energy required for transmission when $D_\mathrm{S} \ge \SI{11}{m}$, which is due to the increased number of transmission attempts that are required for each packet.
Conversely, the trend of the average data rate for the attempts (which depends on Minstrel operation) is similar for both the VISIBLE and HIDDEN configurations, and resembles the NO\_INT configuration (with no interferers).
This implies that also the duration of frames is similar, and confirms that, in the HIDDEN configuration, the increase in power consumption is caused by the higher number of retransmissions and not by the longer transmission times due to lower data rates.

\section{Discussion and Conclusion}
\label{sec:Conclusions}
For many reasons, including performance, dependability, and cost, Wi-Fi is increasingly considered one of the most suitable wireless communication technologies for interconnecting moving devices in industrial environments.
The primary objective of this work is to study some of the most common phenomena that affect the quality of service in industrial Wi-Fi networks and quantify their effects.
Besides reliability and timeliness, which are the traditional metrics of interest in industrial scenarios, sustainability aspects such as energy and spectrum consumption are also taken into account.
Moreover, we explicitly considered (and discussed) the effects of Minstrel, as rate adaptation is quite effective and customarily supported by most commercial \mbox{Wi-Fi} equipment.

Three network configurations of increasing complexity have been analyzed using the ns-3 network simulator.
The first configuration (NO\_INT) foresees a wireless station associated to an access point.
To analyze the effect of signal attenuation, the QoS is evaluated by varying their distance $D_\mathrm{S}$ from $\SI{1}{m}$ to $\SI{50}{m}$.
In the second (VISIBLE) and third (HIDDEN) configurations, an interfering wireless station is introduced as well. 
Two subcases are considered, where the interferer is located close to the AP and at $\SI{40}{m}$ from it (in the opposite direction with respect to the SUT), respectively.

For the configuration with no interferers, which acts as the baseline, results show a latency increase when $D_\mathrm{S} \geq \SI{28}{m}$ due to failed attempts at higher data rates, which is mitigated by Minstrel by lowering the transmission speed.
Higher power consumption is experienced, as more than one attempt is needed on average, and their duration is longer.
Results also highlight that, due to MCS exploration, Minstrel performance is suboptimal.

The shape of latency plots, when the interferer is visible (and collision avoidance operates as intended), resembles the case with no interferers, but values are higher because the channel is found busy more often due to the interfering traffic.
Power consumption does not increase appreciably because no energy is spent while waiting for the channel to become idle again.

Finally, a specific configuration focuses on the effects of the hidden node.
When the SUT and INT are spaced by more than $\SI{51.45}{m}$, they are not able to sense each others' traffic, which increases collisions noticeably.
This results in packets being dropped (up to $0.7\%$, in our experimental configuration) and in a higher number of attempts needed for delivering packets (on average, up to $1.47$ per packet, see Fig.~\ref{fig:AvgLatInt}).
Retransmissions due to collisions cause a latency increase and also impact power consumption significantly.
This is not the same as for the other configurations, where energy consumption was mostly affected by the increase in the frame duration due to Minstrel.

Concerning Wi-Fi suitability for (soft) real-time systems, we note that worst-case latency (pragmatically evaluated as its $99.9$ percentile) in the presence of a hidden interferer is one order of magnitude higher than when it is visible.
As a consequence, the hidden node proves to be a severe issue, since small changes in the position of mobile devices lead to large variations in the QoS they receive for communication.
This may impair the effectiveness of models for estimating the QoS from the (approximate) position and traffic of nodes.

In conclusion, the obtained results are intended as the initial step for developing simple and fast approximate numerical models for the analysis and optimization of industrial \mbox{Wi-Fi}  networks.
In particular, our future research work will focus on the design and development of network digital twins for Wi-Fi,
which rely on the above models to enable execution on devices with (relatively) low computational power, like APs.
Finally, the presented results will be extended to multi-dimensional scenarios (2D/3D) and will also consider to what extent rate adaptation algorithms (that basically rely on low-pass filters) are affected by non-slow SUT movements.
Future activities should also simulate more realistic industrial environments, where the presence of obstacles and surfaces can lead to multi-path propagation and shadowing.

\bibliographystyle{IEEEtran}
\bibliography{bibliography}

\cleardoublepage

\end{document}